\documentclass[11pt]{article}

\usepackage[percent]{overpic}
\usepackage{float}
\usepackage{pgfplots}
\usepackage{placeins}
\usepackage{amsmath}
\usepackage{amsthm}
\usepackage{color}
\usepackage{amssymb}
\usepackage{mathtools}
\usepackage{subfigure}
\usepackage{multirow}
\usepackage{epsfig}
\usepackage{listings}
\usepackage{enumitem}
\usepackage{rotating,tabularx}
\usepackage{graphicx}
\usepackage{graphics}
\usepackage{epstopdf}
\usepackage{longtable}
\usepackage[pdftex]{hyperref}
\usepackage{epigraph}
\usepackage{xspace}
\usepackage{amsfonts}
\usepackage{eurosym}
\usepackage{ulem}
\usepackage{footmisc}
\usepackage{comment}
\usepackage{setspace}
\usepackage{geometry}
\usepackage{caption}
\usepackage{pdflscape}
\usepackage{array}
\usepackage[round]{natbib}
\usepackage{booktabs}
\usepackage{dcolumn}
\usepackage{mathrsfs}
\usepackage{tabularx}
\usepackage{enumitem}
\usepackage{tikz}
\usetikzlibrary{decorations.pathreplacing}

\usepackage{xcolor}
\hypersetup{
colorlinks,
linkcolor={blue!50!black},
citecolor={blue!50!black},
urlcolor={blue!50!black}}

\newcommand{\bc}{\begin{center}}
\newcommand{\ec}{\end{center}}

\newcolumntype{d}[1]{D{.}{.}{#1}} 

\newcolumntype{L}[1]{>{\raggedright\let\newline\\arraybackslash\hspace{0pt}}m{#1}}
\newcolumntype{C}[1]{>{\centering\let\newline\\arraybackslash\hspace{0pt}}m{#1}}
\newcolumntype{R}[1]{>{\raggedleft\let\newline\\arraybackslash\hspace{0pt}}m{#1}}

\newcommand{\Cref}[1]{Corollary~\ref{coro:#1}}
\newcommand{\cref}[1]{Cor.~\ref{coro:#1}}

\newcommand{\be}{\begin{equation}}
\newcommand{\ee}{\end{equation}}
\newcommand{\bea}{\begin{eqnarray}}
\newcommand{\eea}{\end{eqnarray}}
\newcommand{\bi}{\begin{itemize}}
\newcommand{\ei}{\end{itemize}}

\numberwithin{equation}{section}

\usepackage{xcolor}
\usepackage{graphicx}
\usepackage{adjustbox}
\usepackage{lscape} 
\usepackage{array}
\usepackage{longtable}
\begin{document}
\begin{titlepage}
\title{Determinants of budget deficits: Focus on the effects from the COVID-19 crisis}
\author{Dragan Tevdovski \footnote{Faculty of Economics, University Ss. Cyril and Methodius in Skopje,~\url{dragan@eccf.ukim.edu.mk}} \and Petar Jolakoski\footnote{Association for Research and Analysis – ZMAI,~\url{jolakoskip@gmail.com}} \and Viktor Stojkoski\footnote{Faculty of Economics, University Ss. Cyril and Methodius in Skopje and Association for Research and Analysis – ZMAI,~\url{vstojkoski@eccf.ukim.edu.mk}}\,\, }
\date{\today}
\maketitle
\begin{abstract}
This paper revisits the discussion on determinants of budget balances and investigates the change in their effect in light of the Covid-19 crisis by utilizing data on 43 countries and a system generalized method of moments approach. The results show that the overall impact of the global pandemic led to a disproportionate increase in the magnitude of the estimated effects of the macroeconomic determinants on the budget balance. However, we also find that more developed economies were able to undertake higher stimulus packages for the relatively same level of primary balance. We believe that one of the factors affecting this outcome is that that they exhibit a higher government debt position in domestic currency denomination.
\noindent 
\\
\\
\noindent\textbf{budget deficits, economic determinants, COVID-19}
\end{abstract}
\setcounter{page}{0}
\thispagestyle{empty}
\end{titlepage}
\pagebreak \newpage

\section{Introduction}\label{sec:introduction}
\label{sec:intro}

Governments all around the world have taken forceful economic measures as a response to the sharp decline in global output due to the wide scale economic disruption caused by the COVID-19 pandemic. Both fiscal and monetary authorities stepped in to provide immediate liquidity support, replace lost household income, save jobs and prevent large-scale bankruptcies. However, the measures taken are costly. In particular, along with the drastic fall in government tax revenues and the abrupt rise in government expenditures, the measures are expected to push global public debt to an all-time high. In fact, the International Monetary Fund (IMF) estimated that global public debt will increase by 16 percentage points in 2020, from 83\% in 2019 to around 100\% of GDP in 2020, which is unparalleled to any increase in the past. 

The focus of this paper is to review and revisit the determinants of budget balances in light of the global pandemic in 2020. We make an attempt to answer several compelling questions such as:  Has the effect of a rise in government expenditure on the budget balance changed in 2020? How does the state of the labour market in 2020 affect the budget balance? Are budget balances in 2020 constrained by the prevailing debt levels and the current low long-term interest rates? Does the existence of previous vulnerabilities matter for the change in the budget balance in 2020?

To provide the answers, we use a system-GMM estimation procedure where we introduce interaction terms for macroeconomic variables and a dummy variable for 2020 in an otherwise standard specification including macroeconomic, political/institutional and demographic variables, as suggested by the literature~\citep{alesina1998political,agnello2009determinants,maltritz2015determinants}. For this purpose, we utilize a dynamic panel dataset with 43 countries (all countries for which sufficient data is available) from 1995 to 2020. We use actual realizations of all of the series from 1995 to 2019 and IMF WEO forecasts of the series for 2020. The system approach allows us to address and overcome the issue of endogeneity that arises among the independent variables that is a not accounted for in a standard least squares regression. 

We contribute to the literature in several ways. First, a growing body of literature aims to determine the size of the budget deficits in 2020, and offer guidance on policies for sustainable repayment of the increased debt~\citep{makin2021global}. However, to the best of our knowledge, the literature lacks investigations regarding the role of economic determinants on the magnitude of the budget deficits in 2020. This is among the first papers that attempt to quantify and analyze the potential difference in the effect of standard macroeconomic determinants on the budget balance prior to and during the COVID-19 pandemic. In addition, the paper offers a lengthy discussion on the implications of this change in terms of the long term economic scars that might result from the on-going crisis. 

Our results indicate that the overall impact of the global pandemic led to a disproportionate increase in the magnitude of the estimated effects of almost all of the macroeconomic determinants on the budget balance. In particular, the effects of the economic slowdown, the rise in government debt and government expenditure as well as population size on the budget balance are greater in 2020 in comparison to the preceding period. Thus, in 2020 larger public debts imply even lower budget deficits, suggesting that higher debt in the midst of a severe economic downturn constrained additional spending by the government.

We also discuss our findings by studying the relationship between the magnitude of the primary balance in 2020, the fiscal packages introduced to tackle the consequences of the COVID-19 pandemic and the government debt position by currency of denomination. We argue that more developed economies were able to undertake higher stimulus packages for the relatively same level of primary balance, and this was in part due to the fact that they exhibit a higher government debt position in domestic currency denomination (as percent of GDP).

The remainder of the paper is organized as follows: Section~\ref{sec:lit-review} offers a brief overview of the related literature. Section~\ref{sec:methods} introduces the applied methodology and the data used in more detail. Section~\ref{sec:results} provides the empirical results and studies the thereby induced implications. In Section~\ref{sec:discussion} we discuss our findings.

\section{Literature Review}\label{sec:lit-review}

The existing literature studies the budget balance from two main perspectives: economic and political. A majority of the studies that are focused on the economic perspective analyze the response of fiscal policy to output. Standard Keynesian models deliver the need for fiscal policy to be countercyclical: allowing for budget deficits in recessions and saving for budget surpluses in booms, i.e., government spending (taxes) should rise (decrease) in recessions and increase in booms. In fact, the tax smoothing theory of \cite{barro1979determination} argues that the government should smooth both tax rates and government spending by borrowing during recessions and repaying during booms. Put simply, government spending will be uncorrelated with the changes in output, while tax revenue will be positively correlated. So far there is relatively mixed evidence about the cyclicality of fiscal policy in the voluminous literature (for example,\citep{gali2003fiscal,akitoby2004cyclical,talvi2005tax,alesina2008fiscal,frankel2013graduation}). However, the general consensus from these studies is that fiscal policy is counter-cyclical in most developed countries, while it is pro-cyclical in developing countries \citep{afonso2010fiscal}.

Evidently, the explanation for the cross-country variation in fiscal policy cyclicality cannot be provided solely by economic factors. The political characteristics of the governments, their ideological motivations, types of electoral systems and institutional arrangements are also important factors that determine fiscal policy. A large part of the literature focused on political determinants finds that weaker governments in terms of tenure and political power create larger budget deficits. \cite{roubini1989political} showed that countries where governments have short tenures tend to have higher deficits on average. Moreover, the paper shows that multi-party coalition governments have a higher tendency to develop large and persistent deficits in comparison with majority party governments. Similarly, \cite{lane2003cyclical} argued that countries with dispersed political power are most likely to run pro-cyclical fiscal policies. Additionally, higher deficit is found to be positively associated with the size of the government cabinet measured as the number of spending ministers \citep{volkerink2001fragmented,perotti2002fragmented}.

Although it is reasonable to expect that right-wing governments will practice tight fiscal policy and left-wing governments loose fiscal policy, the empirical literature found mixed evidence for the influence of the government ideological preferences on the budget balances \citep{alesina1997political,mulas2003political}. The budget deficits tend to be higher in election years in the case of opportunistic governments that do not have ideological preferences, but just follow policies that maximize their probability to win the next elections \citep{franzese2002electoral,de2005has}. \cite{alesina1995political} found that large deficits are more common in countries with proportional rather than majoritarian and presidential electoral system. In addition, \cite{alesina2008fiscal} argued that most of the pro-cyclicality of fiscal policy in developing countries can be explained by high level of corruption. 

The institutional factors are also positively associated with fiscal performance. \citep{leachman2007political} discussed that fiscal performance is better when fiscal budgeting institutions are strong. \cite{de1997political} found that a strong finance minister or commitment to negotiated budget targets can be especially effective in keeping deficits down in countries where there is some political instability. \cite{henisz2004political} found that checks and balances that limit the discretion of policy‐makers reduce the volatility of government expenditure and revenue. 

Lastly, a relatively new, but important strand of literature is the examination of fiscal persistence. This is a measure of the degree of dependence of current fiscal behavior on its own past developments. In this aspect, \cite{afonso2010fiscal} found that countries with higher fiscal persistence tend to have lower discretion.

\section{Methodology}\label{sec:methods}

\subsection{Model}\label{sec:model}

In the construction of the econometric model we follow the literature described in the previous section and specify it as: 
\begin{equation}
    \texttt{Balance}_{it} = \beta_0 \texttt{Balance}_{it-1} + \mathbf{Y}^{'}_{it} \beta_1 + \mathbf{X}^{'}_{it} \beta_2 + \mathbf{Y}^{'}_{it} D \delta_{\beta_1} + \mathbf{X}^{'}_{it} D \delta_{\beta_2} + v_i + \varepsilon_{it}, 
    \label{eq:model}
\end{equation}
where the dependent variable $\texttt{Balance}_{it}$ is the primary budget balance of country $i$ in time $t$.
We assume that the primary budget balance is dependent on its previous value $\texttt{Balance}_{it-1}$, with $\beta_0$ being its marginal effect and two disjoint sets of explanatory variables $\mathbf{Y}^{'}_{it}$ and $\mathbf{X}^{'}_{it}$ \citet{agnello2009determinants,maltritz2015determinants}. The first set has a marginal effect $\beta_1$ and describes potential macroeconomic determinants of the primary budget balance, whereas the second set is constituted of political and institutional variables, and $\beta_2$ is their marginal effect. In addition, in the equation we include a country specific effect $v_i$ to account for potentially omitted variables that are invariant over time. 

Similar model specifications have been used in~\citep{alesina1998political,agnello2009determinants,maltritz2015determinants} to understand the critical factors that drive the magnitude and characteristics of the budget deficit in an economy. The novelty in our specification is the presence of the interaction term between the independent variables and a dummy variable $D$ for 2020, used for quantifying the potential different effect of the variables due to the coronavirus pandemic. Concretely, $\delta_{\beta_1}$ and $\delta_{\beta_2}$ represent a direct measure of the change in the effect of the, respectively, macroeconomic and political and institutional variables. A significant and negative value of $\delta_{\beta_1}$ ($\delta_{\beta_2}$) implies that the studied variable had an important and larger contribution in the magnitude of the budget deficit in 2020, compared to the usual contribution.

A general problem of this model specification is the presence of endogeneity because of the potential interdependence between the explanatory variables and the budget balance \citep{agnello2009determinants}. This may lead to biased and inefficient parameter estimates. To account for this problem, we resort to a system GMM parameter estimation. The system GMM solves the endogeneity problem via two steps. In the first step, each variable is first-differenced and thus the potential endogeneity due to correlation between the country-specific effects and the explanatory variables is removed~\cite{arellano1991some}. In the second step, the endogeneity between the dependent and explanatory variables is removed by instrumenting the differenced variables with their available lags in levels: the levels of the dependent variable lagged for two or more periods and the levels of the explanatory variables lagged one or more periods~\cite{blundell1998initial}. 

\subsection{Data}

For the purpose of our analysis, we construct an unbalanced panel annual dataset with macroeconomic, political and institutional variables for 43 countries, with yearly data covering the period from 1995 to 2020. We obtain macroeconomic data from the IMF World Economic Outlook (WEO) October 2020 Database and political and institutional data from Databanks International's Cross-National Time-Series Data Archive and Polity IV Database. Our starting point were all of the countries included in the WEO database, but some of the countries were removed from the initial sample due to lack of sufficient data availability in others databases. 

For the period of 1995 to 2019, we use the actual realizations of the chosen variables. For 2020, however, we use forecasts for the macroeconomic variables (IMF) or an assumption that no change has taken place for the institutional and political variables. Although this is a rather strong assumption, our rationale is that changes in these variables require a multitude of legislative and political actions that can rarely be achieved within the span of one year. Our detailed analysis also showed that for all of the countries, it took at least two years for their polity score to change and it remained at the same value for the last few years of our sample. Furthermore, for the entire dataset, the type of regime variable had little to no variation over the years and finally, in terms of the size of cabinet, the change in the number of ministers in the last period is at most 1 to 2 across majority of countries. 

In all of our specifications, the dependent variable is the general government primary budget balance as \% of GDP. We use this measure because it better matches the discretionary decisions of the fiscal authorities in comparison with the overall budget balance. It does not include interest payments for outstanding debt, which is pilled up from the previous period that is not relevant to our study \citep{maltritz2015determinants}. 

The two sets of explanatory variables are based on the relevant existing literature, described in the previous section. We use a standard set of macroeconomic variables in the first set: lagged primary budget balance, government expenditure (log), gross debt (log), interest rate of government debt securities, unemployment rate, GDP growth rate and population (log).  

For the second set of political and institutional explanatory variables we use:  polity scale, type of regime and size of cabinet. \textit{Polity scale} is a variable that evaluates how democratic a country is on a scale from -10 to 10, where the two extremes imply that the country is either fully autocratic (-10) or is fully democratic (10). \textit{Type of regime} is a categorical variable that provides an estimate for the type of government regime in the country: i) civilian, ii) military-civilian and iii) military. \textit{Size of cabinet} quantifies the number of ministers in a government. The list of all used variables, their transformation and data sources are presented in Table~\ref{tab:app-a} of the Appendix. Table~\ref{tab:app-b} in the Appendix presents a list of all countries included in the sample and their values for the macroeconomic variables for 2019 and 2020. 

Table~\ref{tab:mean_descriptives} summarizes the descriptive statistics for all of the variables that are included in our empirical analysis for 2019 and 2020. These statistics suggest that the primary budget deficit across all countries increased on average by 6.1 percentage points in 2020 (-6.4 percent of GDP) in comparison with 2019 (-0.32 percent of GDP), while the gross government debt increased on average by 12.8 percentage points from 2019 (69.3 percent of GDP) to 2020 (82.1 percent of GDP). This extraordinary increase in the global public debt happened at a time of further decline of average interest rates of government securities by almost 1 p.p. from 2019 (3.4 percent) to 2020 (2.5 percent). During the 2020 crisis, as expected, the loose fiscal policy leads government expenditures to increase on average by 8.7 percentage points, whereas for the unemployment rate the average increase was 2.5 percentage points and for GDP growth rate an average decrease of 8.6 percentage points across the entire sample. These numbers reflect the current economic conditions and the economic support packages implemented by policymakers around the world as a response to the global pandemic. 

The polity scale variable shows that democracy is on a higher level in advanced economies than in emerging economies (9.1 and 5.3, respectively). The average number of ministers in the governments of both country groups is almost the same (around 21). Most of the governments of the countries in our sample are classified as civilian with a few exceptions (Algeria, Egypt, Fiji, Sudan and Thailand classified as military-civilian; and Pakistan and Thailand have a military regime for a relatively short period of time).


\begin{table}[]
\centering
\begin{tabular}{|l|c|c|}
\hline
                        & \multicolumn{2}{c|}{All countries}                                                                                \\ \hline
\textbf{Variable}       & 2019                                                    & 2020                                                    \\ \hline
\multicolumn{3}{|c|}{\textbf{Dependent variable}}                                                                                           \\ \hline
Primary balance         & \begin{tabular}[c]{@{}c@{}}-0.32\\  (2.85)\end{tabular} & \begin{tabular}[c]{@{}c@{}}-6.44\\  (4.24)\end{tabular} \\ \hline
\multicolumn{3}{|c|}{\textbf{Macroeconomic and demographic   variables}}                                                                                    \\ \hline
Gov. gross debt         & \begin{tabular}[c]{@{}c@{}}69.27\\ (38.48)\end{tabular} & \begin{tabular}[c]{@{}c@{}}82.14\\ (46.78)\end{tabular} \\ 
Population              & \begin{tabular}[c]{@{}c@{}}46.42\\ (73.33)\end{tabular} & \begin{tabular}[c]{@{}c@{}}47.2\\ (77.73)\end{tabular}  \\ 
Unemployment   rate     & \begin{tabular}[c]{@{}c@{}}6.78\\ (6.37)\end{tabular}   & \begin{tabular}[c]{@{}c@{}}9.24\\ (8.31)\end{tabular}   \\ 
Interest rate                   & \begin{tabular}[c]{@{}c@{}}3.41\\ (3.43)\end{tabular}   & \begin{tabular}[c]{@{}c@{}}2.52\\ (3.3)\end{tabular}    \\ 
Gov. exp. (\%   of GDP) & \begin{tabular}[c]{@{}c@{}}34.77\\ (9.82)\end{tabular}  & \begin{tabular}[c]{@{}c@{}}42.88\\ (11.45)\end{tabular} \\ 
GDP growth   rate       & \begin{tabular}[c]{@{}c@{}}2.47\\ (1.68)\end{tabular}   & \begin{tabular}[c]{@{}c@{}}-6.11\\ (2.80)\end{tabular}  \\ \hline
\multicolumn{3}{|c|}{\textbf{Political and institutional variables}}                                                                        \\ \hline
Polity                  & \begin{tabular}[c]{@{}c@{}}7.57\\ (4.08)\end{tabular}   & \begin{tabular}[c]{@{}c@{}}7.4\\ (4.37)\end{tabular}    \\ 
Size of   cabinet       & \begin{tabular}[c]{@{}c@{}}19.57\\ (5.51)\end{tabular}  & \begin{tabular}[c]{@{}c@{}}19\\ (5.47)\end{tabular}     \\ 
Civilian regime          & 25                                                      & 21                                                      \\ 
Military-civilian regime & 1                                                       & 1                                                       \\ 
Military regime          & 0                                                       & 0                                                       \\ \hline
\end{tabular}
\caption{\textbf{Summary statistics. Mean values per country groups for 2019 and 2020. Standard deviations in brackets.
Note: Regime statistics refer to number of countries}}
\label{tab:mean_descriptives}

\end{table}

In fact, discretionary budget measures on both the spending and the revenue side taken to combat the virus by emerging and developing economies account for more than 3.5\% of GDP and more than 9\% of GDP in advanced economies (IMF WEO October 2020). Given the specific nature of this shock, the severe weakening of aggregate demand and continuous disruptions of aggregate supply are expected to lead to the deepest global recession since World War II (World Bank Global Economic Prospects 2020). As a result, governments across the globe stepped in with extensive fiscal packages along with complimentary institutions to help the ailing economies, that include wage subsidies, tax deferrals, easing of regulatory burdens, transfers to firms and households, postponement of loan repayments, government guarantees among many other measures. Although the economic support offered by all of the countries is unprecedented and higher than what was offered during the global financial crisis in 2009 (World Bank Global Economic Prospects), the state of the economies prior to the emergence of the viral outbreak is also a crucial determinant of the magnitude of their response.

\section{Empirical results}\label{sec:results}

In this section, we discuss the empirical results obtained using the \cite{arellano1991some} and \cite{blundell1998initial} methods of implementing a dynamic linear GMM estimation. A summary of our main findings can be found in Table~\ref{tab:results}. In column 1, we present the results from our model with the primary budget balance and the set of macroeconomic variables. We add additional explanatory variables that have been used in the literature that include a demographic effect (column 2) and a potential effect coming from political/institutional variables (column 3). Finally, in column 4 we present the results that include interaction terms with the variables that we consider to have been affected by the Covid-19 crisis and a dummy variable for 2020.

All of the macroeconomic variables are significant in all of the regressions we estimated and have the expected sign, as typically found in the literature \citep{roubini1989political,bayar2009economic,maltritz2015determinants}. The effect of the lagged primary budget balance is positive and significant. In fact, this persistence in the effect of the primary budget balance corresponds to a well-documented inertia in the budgetary process found in the literature. Similarly, the effect of the GDP growth rate is also positive and significant, as expected, because a higher growth rate of the economy improves the primary budgetary balance on the short run.

Next, government expenditure has a significant and negative impact on the primary budgetary balance as, by construction, higher government expenditure raises the primary budget deficit in the short run. In terms of the unemployment rate, the coefficient is negative and significant, implying that an increase in the unemployment rate worsens the primary budgetary balance, as a result of the additional government expenditure incurred to support the labour market. Another negative impact on the budget balance comes from population (significant in column 3 and 4). 

The stock of debt and the long term interest rate have a significant and positive relationship with the primary budgetary balance. The fact that gross government debt has a positive effect on the primary budget balance confirms a previous finding in the empirical literature that higher debt improves the primary budget balance and reduces deficits. \cite{maltritz2015determinants} argue that high debt implies less fiscal space to encourage additional spending, while low debt levels enable countries to run higher deficits. We underline that higher interest payments from debt do not affect the primary balance which, by definition, excludes interest payments. However, higher long term interest rates on debt instruments also implies less fiscal space and leads to improved budgetary balances. On the other side, lower or negative long term interest rates on debt instruments encourage budget deficits. 

In terms of the set of political/institutional variables, some of the results agree with the empirical findings in the literature. The coefficient in front of the polity scale variable is significant and positive, implying that a more democratic regime tends to be characterized with stronger institutions and functional check and balances that limit the discretion of policy makers to increase budget deficits. This finding is in the line with the literature (for example see \citet{henisz2004political} and \citet{leachman2007political}).
The effect of the size of cabinet is negative but insignificant. A negative impact fits the stylized fact in the literature that higher number of spending ministers is associated with a lower primary balance (for example, \cite{volkerink2001fragmented} and \cite{perotti2002fragmented}).
Finally, in terms of the type of regime, the results show that countries that have a military-civilian or civilian regime tend to have higher primary budget deficits on average. 

The last column shows the change in the effect of the macroeconomic variables on the primary budget balance in the year of the global pandemic. The pandemic increases the magnitude of the estimated effects of all of the variables, except of the unemployment rate where the direction of the relationship is reversed. Overall, the sharp decline in economic activity and increased government expenditures result in a higher budget deficit. This is what we expect, since given the economic toll of the crisis, all of the countries in our sample have implemented some form of economic support packages, resulting in a deterioration of their fiscal health. The resulting rise in the public debt leads to an increased magnitude of the positive effect of gross debt on the primary balance in 2020. This can be an indication that higher debt in the midst of a severe economic downturn implies enhanced incentives against spending. Moreover, population is expected to contribute more negatively to the primary balance in 2020, as with the on-going pandemic, more health related spending is needed in countries with a larger population. The only unexpected result is the positive effect coming from the unemployment rate in 2020, which may suggest that labor market deterioration led to lower government revenue mobilization and further constraints on the primary budget balance.

The robustness of our results is confirmed by re-estimating the model when the interest rate is removed from the list of explanatory variables. We explicitly choose this variable since, when excluded, the sample increases by the largest margin. The results are given in Table~\ref{tab:appresultsnoint} in the Appendix.

\begin{table}[t!]
\resizebox{\textwidth}{!}{\begin{tabular}{lcccc}
 & (1) & (2) & (3) & (4) \\
VARIABLES & Macroeconomic & Demographic & Political/Institutional & 2020 \\ \hline
Primary balance (t-1) & 0.404*** & 0.399*** & 0.374*** & 0.357***\\
 & (0.018) & (0.026) & (0.020) & (0.036) \\
GDP growth rate & 0.244*** & 0.251*** & 0.223*** & 0.173***\\
 & (0.013) & (0.011) & (0.015) & (0.019) \\
Unemployment rate & -0.064*** & -0.067*** & -0.081*** & -0.100* \\
 & (0.022) & (0.019) & (0.030) & (0.053) \\
Government debt & 2.828*** & 2.513*** & 2.830*** & 4.074*** \\
 & (0.376) & (0.319) & (0.337) & (0.656) \\
Government expenditure & -17.267*** & -17.070*** & -19.745*** & -17.660*** \\
 & (0.839) & (1.074) & (1.709) & (1.660) \\
Interest rate & 0.263*** & 0.243*** & 0.198*** & 0.181*** \\
& (0.025) & (0.019) & (0.026) & (0.051) \\
Population &  & -0.219 & -6.323** & -5.924* \\
& & (1.707) & (2.994) & (3.233) \\
Polity & & & 1.052*** & 0.705** \\
& & & (0.250) & (0.308) \\
Size of cabinet & & & -0.032 & -0.027\\
 & &  & (0.031) & (0.040) \\
Type of regime (2) & & & 8.901*** &  5.687 \\
 & & & (1.318) &  (1.618) \\
Type of regime (3) & & & -5.435*** &  2.506 \\
 & & & (3.139) & (3.742) \\
GDP*2020 & & & & 0.984***  \\
 & & & &  (0.241) \\
Unemployment*2020 & & & &  0.647** \\
 & & & & (0.318)\\
Government expenditure*2020 & & & & -5.580*** \\
 & & & & (2.071) \\
Debt*2020 & & & & 6.028*** \\
 & & & & (1.857) \\
Population*2020 & & & & -0.995** \\
 & & & & (0.433) \\
Constant & 49.516*** & 50.961*** & 67.588*** & 58.025*** \\
 & (2.655) & (5.614) & (10.078) & (9.028) \\ \hline
Observations & 854 & 854 & 854 & 854 \\
Number of countries & 43 & 43 & 43 & 43 \\ \hline
Standard errors in parentheses & \multicolumn{1}{l}{} & \multicolumn{1}{l}{} & \multicolumn{1}{l}{} \\
*** p\textless{}0.01, ** p\textless{}0.05, * p\textless{}0.1 & \multicolumn{1}{l}{} & \multicolumn{1}{l}{} & \multicolumn{1}{l}{}
\end{tabular}}
\caption{\textbf{Results. Arellano–Bover/Blundell–Bond linear dynamic panel-data estimation}}\label{tab:results}
\end{table}

\pagebreak

\subsection{Implications}

In the period that follows ahead, economies across the globe are expected to bear unprecedented economic and social costs from the COVID-19 pandemic. Returning to pre-crisis levels of economic activity is a daunting task for countries all over the world and the path to recovery is not to be smooth, even or certain. 

\begin{figure}
\begin{center}
\includegraphics[width=0.7\textwidth]{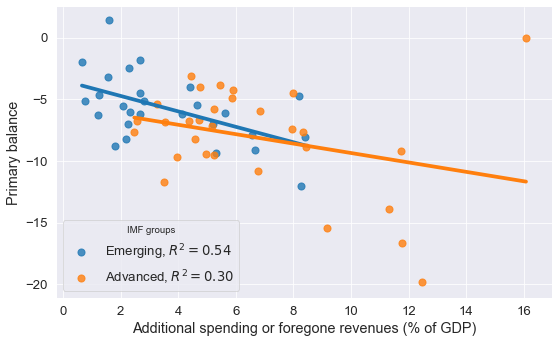}
\end{center}
\caption{\textbf{IMF country groups: Explained variation in budget deficits due to additional spending or foregone revenues in response to COVID-19 pandemic. \\
Source: Fiscal Monitor, Database of Country Fiscal Measures in Response to the COVID-19 Pandemic, IMF.\label{fig:fiscal_measures_wb}}}
\end{figure}

To begin with, the fiscal stimulus packages implemented by governments in response to the COVID-19 put a significant strain on public finances. The unparalleled fiscal response to demand slumps and supply interruptions that followed the crisis was to a large extent made by advanced economies and some large emerging market economies. This is because these countries relied on more favorable financing conditions prior to the crisis and retained the ability to borrow at lower interest rates (IMF, WEO 2020 and IMF, Fiscal Monitor 2020). 

To better understand the relationship between the fiscal stimulus packages introduced by the governments and the budget deficits in 2020 in Figure~\ref{fig:fiscal_measures_wb} we plot the primary balance in 2020 as a function of the additional spending or foregone revenues in response to COVID-19. The source of data for countries' fiscal measures in response to pandemic is the Fiscal monitor database of IMF - we also divide the countries into emerging and advanced on the basis of the IMF's classification. 

The figure shows that some advanced economies were able to undertake higher stimulus packages for the relatively same level of the primary balance. The conclusion is similar when looking at the total fiscal measures undertaken, which additionally include liquidity support in the form of equity injections, loans asset purchase or debt assumptions, as depicted in Figure~\ref{fig:fiscal_measures_wb}. The figure highlights that total fiscal measures were in general higher in advanced economies in comparison with emerging market economies. 

\begin{figure}
\begin{center}
\includegraphics[width=0.7\textwidth]{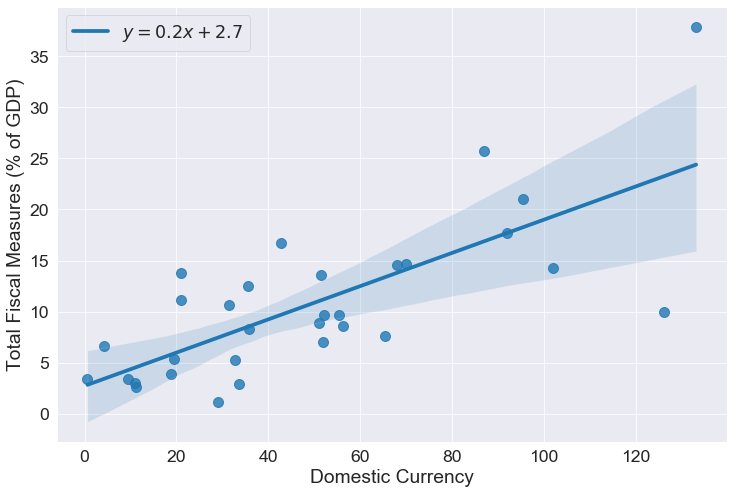}
\end{center}
\caption{\textbf{Explained variation in total fiscal measures due to government debt position in domestic currency denomination (as a percent of GDP) \label{fig:fiscal_measures_domestic_currency}}}
\end{figure}

While country income per capita is an important determinant of the size of total fiscal measures during the pandemic, we want to emphasize that also government debt position in domestic currency denomination was relevant determinant during the COVID-19 crisis. We could not include this variable in our regression model, due to lack of data availability for a lot of the countries in our sample. However, in Table~\ref{tab:app-a} in the Appendix we present all of the countries for which data is available in the Fiscal Monitor of the IMF database and the Quarterly public sector debt database of the World Bank for gross central government debt position by currency of denomination (as a percent of GDP) in the first quarter of 2020. We illustrate the relationship using these countries in Figure~\ref{fig:fiscal_measures_domestic_currency}. The figure shows that countries with a higher government debt position in domestic currency denomination (as percent of GDP) were able to execute higher total fiscal measures (as percent of GDP) during the pandemic. For example, four of five countries with one of the highest total fiscal measures as percent of GDP are Italy, United Kingdom, France and Spain, which have a debt position in domestic currency higher than 80 percent. Also, we note that the correlation coefficient between debt position in domestic currency denomination and size of the total fiscal measures is fairly high, equal to 0.737, and significant on the 1\% level.

\section{Discussion} \label{sec:discussion}

The fiscal packages aimed to stimulate the economy amidst the COVID-19 pandemic led to an unprecedented growth in the 
budget balance deficits in almost every country in the world. Here, we discovered that this growth was accelerated by the change in the impact of the economic determinants that govern the magnitude of the budget deficits. We also postulated that more developed economies were able to undertake higher stimulus for the same level of budget deficits, mainly because of the advantage of servicing the debt in their national currency.

While the fiscal packages had a vital role in the efforts made by governments stepping in to combat the consequences of the virus, it is clear that future fiscal space will be limited as a result of these efforts. The rise in the global debt to GDP ratio can pose additional challenges for debt sustainability of all of the economies in the world in the medium to long run. Equally relevant is the potential risk in the horizon for debt financing conditions. This is because the decades long downward trend of long-term interest rates could be reversed, given that in many countries nominal interest rates are close to their lower bounds.

Moreover, some of the vulnerabilities that existed prior to the crisis, such as population aging, are likely to further contribute negatively to the outlook for the stock of sovereign debt. For instance, in their empirical study on OECD economies, ~\citet{honda2020would} found that population aging weakens fiscal spending effects and in order to support the economy in a downturn, countries will need to revert to larger fiscal support packages. Finally, the buildup of debt is also expected to constrain the use of government revenues for growth and development related spending needs in the future, as a large part of the inflows to the government budget will be consumed by debt service.

Last but not least, the size, distribution and adjustment of the budget balance in 2020 will determine both the general social prospects and industry-wise economic capabilities of each country in the aftermath of the pandemic~\citep{stojkoski2020socio,stojkoski2020short,tevdovski2021impact}. The structural changes induced on the economies as they adjust to constraints imposed by health authorities (social distancing, teleworking, movement impediments, capacity restrictions) will redistribute societal resources from highly inflexible sectors to highly adaptive sectors of the economy. Along that route, welfare losses arising from labour market distortions (jobs at the lower quantiles of the wage distribution, in informal employment, with temporary working arrangements) and human capital accumulation disruptions (schooling interruptions and transformation) will probably worsen the level of poverty and income inequality worldwide. In order to tackle such challenges, mechanisms to restore the fiscal health of governments will be vital in the period to come.


\newpage
\appendix

\section*{Appendix}

\setcounter{equation}{0}
\setcounter{figure}{0}
\setcounter{table}{0}
\setcounter{theorem}{0}
\makeatletter
\renewcommand{\theequation}{A\arabic{equation}}
\renewcommand{\thetable}{A\arabic{table}}
\renewcommand{\thefigure}{A\arabic{figure}}
\renewcommand{\thetheorem}{A\arabic{theorem}}
\renewcommand{\theproposition}{A\arabic{proposition}}

\begin{table}[h]
\caption{\textbf{List of countries and their, fiscal measures, primary budget balance and gross central government debt in domestic currency denomination.} \\ All variables are as \% of GDP.\label{tab:app-a}}
\begin{tabular}{|l|r|r|r|}
\hline
\textbf{Country}         & \textbf{Fiscal measures} & \textbf{Primary balance} & \textbf{ Debt in domestic currency } \\
\hline
Albania         & 2.87                              & -8.42                              & 33.53                                \\
Armenia         & 2.98                              & -5.82                              & 10.90                                 \\
Australia       & 13.53                             & -10.06                             & 51.55                                \\
Brazil          & 14.61                             & -16.78                             & 69.90                                 \\
Bulgaria        & 6.66                              & -2.00                                 & 4.20                                 \\
Canada          & 16.73                             & -19.92                             & 42.69                                \\
Colombia        & 5.27                              & -9.48                              & 32.79                                \\
France          & 20.98                             & -10.77                             & 95.47                                \\
Hungary         & 8.55                              & -8.28                              & 56.17                                \\
Indonesia       & 3.83                              & -6.32                              & 18.92                                \\
Ireland         & 7.59                              & -6.00                                 & 65.48                                \\
Israel          & 9.67                              & -12.94                             & 52.12                                \\
Italy           & 37.93                             & -12.98                             & 133.14                               \\
Lithuania       & 8.28                              & -6.72                              & 35.78                                \\
Luxembourg      & 11.16                             & -6.98                              & 20.89                                \\
Mexico          & 1.10                               & -5.8                               & 28.97                                \\
Moldova         & 2.63                              & -8.00                                 & 11.21                                \\
Netherlands     & 8.85                              & -8.76                              & 51.10                                 \\
Philippines     & 3.37                              & -8.06                              & 0.63                                 \\
Portugal        & 9.95                              & -8.35                              & 126.24                               \\
Romania         & 5.38                              & -9.59                              & 19.51                                \\
Russia          & 3.41                              & -5.29                              & 9.36                                 \\
Slovak Republic & 6.97                              & -8.84                              & 51.82                                \\
Slovenia        & 14.56                             & -8.82                              & 68.01                                \\
South Africa    & 9.61                              & -14.04                             & 55.45                                \\
Spain           & 17.66                             & -14.09                             & 92.06                                \\
Sweden          & 10.63                             & -5.90                               & 31.48                                \\
Thailand        & 12.46                             & -5.21                              & 35.58                                \\
Turkey          & 13.75                             & -7.88                              & 21.10                                 \\
United Kingdom  & 25.72                             & -16.46                             & 86.99                                \\
United States   & 14.22                             & -18.72                             & 101.88  \\            \hline                
\end{tabular}

\end{table}

\begin{table}[t!]
\footnotesize
\caption{List of variables and data sources\label{tab:app-b}}
\renewcommand{\arraystretch}{2}
\resizebox{\textwidth}{!}{\begin{tabular}{ll}
 \hline
 \textbf{Variables}  & \textbf{Source}\\
 \hline
 Primary balance  &  IMF, WEO; General government primary net lending/borrowing \% of GDP \\
 Debt  &   IMF, WEO; General government gross debt \% of GDP\\
 Unemployment rate &  IMF, WEO; \% of total labour force \\
 GDP &   IMF, WEO; Gross domestic product at constant prices\\
 Population  & IMF, WEO in persons (millions) \\
 Government expenditure &  IMF, WEO; \% of GDP\\
 Size of cabinet &  CNTS (polit10) \\
 Type of regime    & CNTS (polit02) \\
 Polity scale &  Polity IV Database (polity2) \\
 \hline
\end{tabular}}
\end{table}

\begin{table}[t!]
\caption{\textbf{Results without the interest rate. Arellano–Bover/Blundell–Bond linear dynamic panel-data estimation}}\label{tab:appresultsnoint}
\renewcommand{\arraystretch}{1.1}
\resizebox{\textwidth}{!}{\begin{tabular}{lcccc}
 & (1) & (2) & (3) & (4) \\
VARIABLES & Macroeconomic & Demographic & Political/Institutional & 2020 \\ \hline
Primary balance (t-1) & 0.522***	&	0.528***	&	0.512***	&	0.536***\\
& (0.003)	&	(0.005)	&	(0.006)	&	(0.006) \\
GDP growth rate & 0.207***	&	0.221***	&	0.213***	&	0.093***\\
 &(0.006)	&	(0.010)	&	(0.010)	&	(0.007) \\
Unemployment rate & -0.025**	&	-0.020	&	-0.050***	&	-0.021 \\
 & (0.010)	&	(0.013)	&	(0.016)	&	(0.019) \\
Government debt & 0.332***	&	0.279***	&	0.039	&	0.470*** \\
 & (0.095)	&	(0.050)	&	(0.080)	&	(0.128) \\
Government expenditure & -13.109***	&	-12.991***	&	-13.530***	&	-12.511*** \\
 & (0.325)	&	(0.395)	&	(0.478)	&	(0.511) \\
Population &	&	0.090	&	0.497***	&	0.631*** \\
&	&	(0.089)	&	(0.140)	&	(0.192) \\
Polity &	&	&		0.118***	&	0.103***\\
&	&	&		(0.018)	&	(0.033) \\
Size of cabinet & & & -0.053***	&	-0.034***\\
 & &  & (0.009)	&	(0.007) \\
Type of regime (2) & & & -1.008	&	-0.721 \\
 & & & (0.699)	&	(1.304) \\
Type of regime (3) & & & -1.395	&	-4.453** \\
 & & & (2.218)	&	(1.873) \\
GDP*2020 & & & & 0.140***  \\
 & & & &  (0.048) \\
Unemployment*2020 & & & &  0.096 \\
 & & & & (0.098)\\
Government expenditure*2020 & & & & -2.948*** \\
 & & & & (0.271) \\
Debt*2020 & & & & 1.966*** \\
 & & & & (0.279) \\
Population*2020 & & & & -0.514*** \\
 & & & & (0.127) \\
Constant & 43.636***	&	43.094***	&	45.553***	&	40.123*** \\
 & (1.320)	&	(1.490)	&	(1.735)	&	(1.991) \\ \hline
Observations & 1,879	&	1,879	&	1,879	&	1,879 \\
Number of countries & 87	&	87	&	87	&	87 \\ \hline
Standard errors in parentheses & \multicolumn{1}{l}{} & \multicolumn{1}{l}{} & \multicolumn{1}{l}{} \\
*** p\textless{}0.01, ** p\textless{}0.05, * p\textless{}0.1 & \multicolumn{1}{l}{} & \multicolumn{1}{l}{} & \multicolumn{1}{l}{}
\end{tabular}}
\end{table}
\pagebreak
\newpage

\end{document}